\documentclass[aps,prl,reprint,preprintnumbers,floatfix,twocolumn,superscriptaddress,amsmath,showpacs,flushbottom,a4paper]{revtex4-1}
\usepackage{amssymb,dsfont}
\usepackage{graphicx}
\begin{document}
\title{Strangeness Contribution to the Proton Spin from Lattice QCD} 
\author{Gunnar S.~Bali}\email{gunnar.bali@ur.de}
\affiliation{Institut f\"ur Theoretische Physik, Universit\"at Regensburg,
              93040 Regensburg, Germany}
\author{Sara~Collins}\email{sara.collins@ur.de}
\affiliation{Institut f\"ur Theoretische Physik, Universit\"at Regensburg,
              93040 Regensburg, Germany}
\author{Meinulf~G\"{o}ckeler}        
\affiliation{Institut f\"ur Theoretische Physik, Universit\"at Regensburg,
              93040 Regensburg, Germany}
\author{Roger~Horsley}  
\affiliation{School of Physics, University of Edinburgh, Edinburgh EH9 3JZ,
              United Kingdom}                                       
\author{Yoshifumi~Nakamura}  
\affiliation{RIKEN Advanced Institute for Computational Science, Kobe, Hyogo 650-0047, Japan}
\author{Andrea~Nobile}  
\affiliation{JSC, Research Center J\"{u}lich, 52425 J\"{u}lich, Germany}
\author{Dirk~Pleiter}  
\affiliation{JSC, Research Center J\"{u}lich, 52425 J\"{u}lich, Germany}
\affiliation{Institut f\"ur Theoretische Physik, Universit\"at Regensburg,
              93040 Regensburg, Germany} 
\author{P.E.L.~Rakow}  
\affiliation{Theoretical Physics Division, Department of Mathematical Sciences,
              University of Liverpool, Liverpool L69 3BX, United Kingdom}
\author{Andreas Sch\"{a}fer}  
\affiliation{Institut f\"ur Theoretische Physik, Universit\"at Regensburg,
              93040 Regensburg, Germany}  
\author{Gerrit Schierholz}  
\affiliation{Deutsches Elektronen-Synchrotron DESY, 22603 Hamburg, Germany}
\author{James M.~Zanotti} 
\affiliation{CSSM, School of Chemistry \& Physics, University of Adelaide,
Adelaide SA 5005, Australia}

\collaboration{QCDSF Collaboration}
\date{\today}
\begin{abstract}
We compute the strangeness and light-quark contributions
$\Delta s$, $\Delta u$ and $\Delta d$ to the proton spin
in $n_{\mathrm{f}}=2$ lattice QCD at a pion mass of about
285~MeV and at a lattice spacing $a\approx 0.073$~fm,
using the non-perturbatively improved
Sheikholeslami-Wohlert Wilson action. We carry out
the renormalization of these matrix elements which
involves mixing between contributions from different quark flavours.
Our main result is the small negative value
$\Delta s^{\overline{\mathrm{MS}}}(\sqrt{7.4}\,\mathrm{GeV})=-0.020(10)(4)$
of the strangeness contribution to the
nucleon spin. The second error is an estimate of the uncertainty,
due to the missing extrapolation to the physical point.
\end{abstract}
\preprint{Adelaide ADP-11-43/T765, Edinburgh 2011/39, Liverpool LTH 934}
\maketitle
\textit{Introduction.}---The proton spin can be split
into a quark spin contribution
$\Delta\Sigma$, a quark angular
momentum contribution $L_q$ and a gluonic contribution
$\Delta G$ (including spin and angular momentum)~\cite{hep-ph/9603249}:
\begin{equation}
\frac12=\frac12\Delta\Sigma+L_q+\Delta G\,.
\end{equation}
In the na\"{\i}ve non-relativistic SU(6) quark model, $\Delta \Sigma = 1$,
with vanishing $L_q$ and $\Delta G$.
In this case there will be no strangeness contribution
$\Delta s$ to
$\Delta \Sigma = \Delta u+\Delta d+\Delta s +\cdots$,
where, in our notation, $\Delta q=\Delta\Sigma_q$  contains both, the spin of the quarks
$q$ and of the antiquarks $\bar{q}$.

Experimentally, 
$\Delta s$ is obtained by integrating the strangeness contribution
$\Delta s(x)$ to the spin structure function $g_1$ over
the momentum fraction $x$.
The integral over the range in which data exist
agrees with zero; see, e.g., new COMPASS
data~\cite{arXiv:1001.4654,arXiv:1007.4061} for $x\geq 0.004$
or HERMES data~\cite{hep-ex/0609039} for $x\geq 0.02$, while
global analyses give
values~\cite{arXiv:1010.0574,arXiv:1007.0351,arXiv:0904.3821}
$\Delta s\approx -0.12$,
suggesting a large negative $\Delta s(x)$ at very small $x$.
Pioneering lattice simulations
of disconnected matrix elements also indicated
values~\cite{hep-ph/9502334,498992} $\Delta s\approx -0.12$.
However, the errors given in these studies are quite
optimistic
while the global fits rely on an extrapolation
of the integrated experimental $\Delta\Sigma$ to small $x$ and constrain
the axial octet charge $a_8$ to a value, obtained
from hyperon $\beta$-decays, assuming
SU(3)$_F$ flavour symmetry. 
Some time ago, employing heavy baryon chiral perturbation theory,
Savage and Walden~\cite{hep-ph/9611210} pointed out that SU(3)$_F$ symmetry
in weak baryonic decays may be violated by as much as 25~\%
and hence $\Delta s(x)$ could remain close to zero also for
$x<0.001$; see also~\cite{Bass:2009ed}.
SU(3)$_F$ symmetry is however supported by
lattice simulations of hyperon axial
couplings~\cite{hep-lat/0208017,arXiv:0712.1214,arXiv:0911.2447,arXiv:1102.3407}, albeit within non-negligible errors. 

In this Letter, we directly compute the matrix
elements that contribute to
the $\Delta q$, including quark line disconnected
diagrams. Preliminary results were
presented at conferences~\cite{arXiv:1112.0024,arXiv:1011.2194,arXiv:0911.2407}.

\textit{Simulation details and methods.}---We simulate $n_{\mathrm{f}}=2$ non-perturbatively
improved Sheikholeslami-Wohlert Fermions, using the
Wilson gauge action, at $\beta=5.29$
and $\kappa=\kappa_{ud}=0.13632$. 
Setting the scale from the chirally extrapolated nucleon mass~\cite{sternbeck},
we obtain the lattice spacing
$a^{-1}=2.71(2)(7)\,\mathrm{GeV}$,
where the errors are statistical and from the extrapolation,
respectively.

We realize two additional
valence $\kappa$ values,
$\kappa_m=0.13609$ and $\kappa_s=0.13550$. The corresponding
pion masses are
$m_{\mathrm{PS},ud}=285(3)(7)\,\mathrm{MeV}$,
$m_{\mathrm{PS},m}=449(3)(11)\,\mathrm{MeV}$ and
$m_{\mathrm{PS},s}=720(5)(18)\,\mathrm{MeV}$.
$\kappa_s$ was fixed so that the $m_{\mathrm{PS},s}$ value
is close to the mass of
a hypothetical strange-antistrange pseudoscalar meson:
$(m_{K^{\pm}}^2+m_{K^0}^2-m_{\pi^{\pm}}^2)^{1/2}\approx
686.9$~MeV.
We investigate volumes of
$32^364$ and $40^364$ lattice points,
i.e., $Lm_{\mathrm{PS},ud}= 3.36$ and 4.20,
respectively, where the largest spatial lattice extent is
$L\approx 2.91$~fm.

The quark polarizations are extracted from the large-time
behaviour of ratios of three-point over
two-point functions. We create a polarized proton at a time $t_0=0$,
probe it with an axial current at a time $t$ and destroy the
zero momentum proton at $t_{\mathrm{f}}>t>0$. 
Quark line connected and disconnected terms contribute:
\begin{align}
R^{\mathrm{con}}(t_{\mathrm{f}},t) &=  \frac{\langle\Gamma_{\rm pol}^{\alpha\beta} C^{\beta\alpha}_{3pt}(t_{\mathrm{f}},t) \rangle}{\langle \Gamma_{\rm unpol}^{\alpha\beta}C^{\beta\alpha}_{2pt}(t_{\mathrm{f}}) \rangle} \,,\\
R^{\mathrm{dis}}(t_{\mathrm{f}},t) &=  -\frac{\langle \Gamma_{\rm pol}^{\alpha\beta}C^{\beta\alpha}_{2pt}(t_{\mathrm{f}}) \sum_{\mathbf{x}}\mathrm{Tr}[\gamma_j\gamma_5 M^{-1}(\mathbf{x},t;\mathbf{x},t)]\rangle}{\langle \Gamma_{\rm unpol}^{\alpha\beta} C^{\beta\alpha}_{2pt}(t_{\mathrm{f}})\rangle}\,.\nonumber
\end{align}
Here $M$ is the lattice Dirac operator,
$\Gamma_{\mathrm{unpol}}=\frac12(\mathds{1}+\gamma_4)$ is a parity projector
and $\Gamma_{\mathrm{pol}}=i\gamma_j\gamma_5\Gamma_{\mathrm{unpol}}$
projects out the difference between the two polarizations
(in direction $\hat{\boldsymbol{\jmath}})$. We average over
$j=1,2,3$ to increase statistics.
For the up and down quark matrix elements we
compute the sum of connected and disconnected terms
while only $R^{\mathrm{dis}}$ contributes to $\Delta s$.

\begin{figure}
\centerline{\includegraphics[height=.44\textwidth,angle=270,clip=]{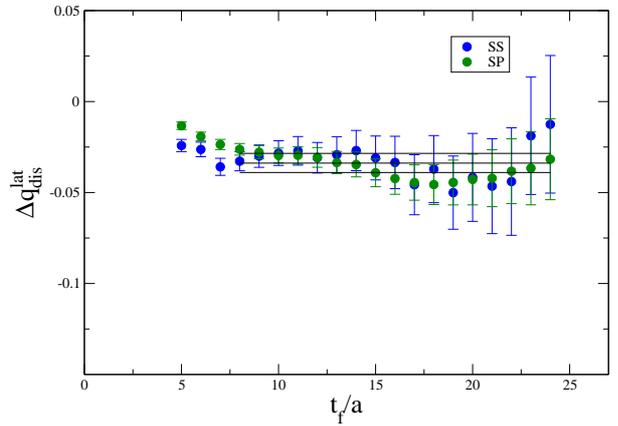}}
\caption{The disconnected ratio $R^{\mathrm{dis}}$ versus $t_{\mathrm{f}}$ 
on the $40^364$ volume
at $\kappa_{\mathrm{val}}=\kappa_{\mathrm{cur}}=\kappa_s$
for smeared-smeared (SS) and smeared-point (SP) source-sink combinations.
\label{fig:ratio2}}
\end{figure}

For disconnected contributions we fix the
time distance between the source and the current
insertion $t=4a\approx 0.29$~fm and vary $t_{\mathrm{f}}$.
Both $t$ and the distance between current
and sink $t_{\mathrm{f}}-t$ should be taken large, to suppress
excited state contributions. 
Using the sink and source smearing
described in~\cite{arXiv:1111.1600},
we find the asymptotic limit
to be effectively reached for $t_{\mathrm{f}}\simeq 6a$--$7a$;
see Fig.~\ref{fig:ratio2} for an example.
The saturation into a plateau at $t_{\mathrm{f}}\leq 2t$ and 
the convergence of the point sink data towards the same value
demonstrate that $t=4a$ was reasonably chosen.
To be on the safe side, we only fit the $t_{\mathrm{f}}\geq 8a\approx 0.58$~fm
smeared-smeared ratios.
Building upon previous experience~\cite{Khan:2006de},
the connected part, for which the statistical accuracy is less of an
issue, is obtained
at the larger, fixed value $t_{\mathrm{f}}=15a$, varying $t$.

The disconnected contribution is computed with the stochastic
estimator  methods described
in~\cite{arXiv:0910.3970,arXiv:1011.2194}, employing time
partitioning, a second order hopping parameter expansion
and the truncated solver method.
We compute the Green functions for four equidistant
source times on each gauge configuration.
We also construct backwardly propagating nucleons, replacing
the positive parity
projector $\frac12(\mathds{1}+\gamma_4)$ by
$\frac12(\mathds{1}-\gamma_4)$, seeding the noise vectors
on eight (four times two) timeslices.
In addition to the 48 (four times spin times colour) solves for
smeared conventional sources, that are necessary
to construct the two-point functions, we run the Conjugate Gradient (CG)
algorithm on $N_1=730$ complex $\mathds{Z}_2$ noise sources for $n_{\mathrm{t}}=40$ iterations.
The  bias from this truncation is corrected for~\cite{arXiv:0910.3970} by
$N_2=50$ BiCGstab solves that are run to convergence.
We analyse a total of 2024 thermalized trajectories on
each of the two volumes where we bin the data to eliminate autocorrelations.

\textit{Renormalization.}---Non-singlet axial currents renormalize with a
renormalization factor $Z_A^{\mathrm{ns}}(a)$ that only depends on the
lattice spacing.
This was determined
non-perturbatively for the action and lattice spacing
in use~\cite{Gockeler:2010yr}: $Z_A^{\mathrm{ns}}=0.76485(64)(73)$.

However, due to the axial anomaly, the renormalization
constant of singlet currents, $Z_A^{\mathrm{s}}(\mu,a)$, acquires an
anomalous dimension. To first
non-trivial order this reads~\cite{Kodaira:1979pa,Larin:1993tq}
$\gamma_A^{\mathrm{s}}(\alpha_s)=-6C_{\mathrm{F}}n_{\mathrm{f}}[\alpha_s/(4\pi)]^2$.
$Z_A^{\mathrm{s}}$ deviates from $Z_A^{\mathrm{ns}}$ starting at $\mathcal{O}(\alpha_s^2)$
in perturbation theory. Both factors have been calculated
to this order, with the result for the conversion
into the $\overline{\mathrm{MS}}$ scheme at a scale $\mu$~\cite{Skouroupathis:2008mf}
\begin{align}
z(\mu,a)&=Z_A^{\mathrm{s}}(\mu,a)-Z_A^{\mathrm{ns}}(a)\nonumber\\
&=C_{\mathrm{F}}n_{\mathrm{f}}\left[15.8380(8)-6\ln(a^2\mu^2)\right]
\left(\frac{\alpha_s}{4\pi}\right)^2\,,
\end{align}
where we have set the Sheikholeslami-Wohlert parameter $c_{\mathrm{SW}}=1$
to be consistent to this order in
perturbation theory.
To this first non-trivial order no scale enters the coupling
parameter $\alpha_s$.
Since perturbation theory in terms of the bare lattice parameter
$\alpha_0=6/(4\pi \beta)$ is known to converge poorly,
we substitute $\alpha_s$
by a coupling defined
from the measured average plaquette 
$\alpha_s=-3\ln \langle U_{\Box}\rangle/(4\pi)=0.14278(5)$,
where we have used the chirally extrapolated value~\cite{Gockeler:2005rv}
$\langle U_{\Box}\rangle=0.54988(11)$.

No dimension-four operator can be constructed that
mixes with the relevant forward matrix element of
$\bar{q}\gamma_{\mu}\gamma_5q$ and that cannot be
removed, using the equations
of motion~\cite{Capitani:2000xi}.
This also holds for the singlet case~\cite{hep-lat/0511014},
such that
we only need to replace
\begin{equation}\label{eq:dqa}
Z_A^{\mathrm{ns}}\mapsto Z_A^{\mathrm{ns}}(1+b_Aam)\,,\quad
Z_A^{\mathrm{s}}\mapsto Z_A^{\mathrm{s}}(1+b^{\mathrm{s}}_Aam)\,,\,
\end{equation}
to achieve full $\mathcal{O}(a)$ improvement.
The factor $b_A$ is known to $\mathcal{O}(\alpha_s)$~\cite{Capitani:2000xi}:
$b_A=b_A^{\mathrm{s}}+\mathcal{O}(\alpha_s^2)\approx 1+18.02539\, C_{\mathrm{F}}\frac{\alpha_s}{4\pi}$.
We obtain the values
\begin{equation}
1+b_Aam=\left\{
\begin{array}{lr}
1.0324(3)(47)&(m_s\,,\kappa=0.13550)\\
1.0041(3) (5)&(m_{ud}\,,\kappa=0.13632)\\\end{array}\,,\right.
\end{equation}
where the first error is due to the uncertainty in the quark mass and
the second error corresponds to 50\,\% of the one-loop correction.
Considering the small size of this correction
it is unlikely that the (two-loop) difference between singlet
and non-singlet $b_A$-factors will result in any noticeable effect,
and in particular not at the light-quark mass $m_{ud}$, where it
will be needed [see Eq.~(\ref{eq:reno}) below].

For $n_{\mathrm{f}}=2$ we get
\begin{equation}
z(\sqrt{7.4}\,\mbox{GeV})=0.0055(1)(27)\,,
\end{equation}
at the renormalization scale 
$\mu^2=7.4\,\mathrm{GeV}^2=1.01(5)\,a^{-2}$.
We again include a 50~\% systematic error to allow for
higher order corrections.
Due to the small anomalous dimension that only sets in at
$\mathcal{O}(\alpha_s^2)$, the difference between singlet
and non-singlet renormalization
constants remains small, also at other scales.
For instance, we obtain $z(\sqrt{10}\,\mbox{GeV})=0.0049(25)$
and $z(2\,\mbox{GeV})=0.0082(41)$.

In the $n_{\mathrm{f}}=1+1+1$ theory the matrix elements
renormalize as follows:
\begin{align}
g_A=\Delta T_3&=(\Delta u -\Delta d )^{\overline{\mathrm{MS}}}\nonumber\\\label{eq:dq3}
&=Z^{\mathrm{ns}}_A(a)(\Delta u -\Delta d)^{\mathrm{lat}}(a)\,,\\
a_8=\Delta T_8&=(\Delta u +\Delta d -2\Delta s)^{\overline{\mathrm{MS}}}\nonumber\\\label{eq:dq2}
&=Z^{\mathrm{ns}}_A(a)(\Delta u +\Delta d -2\Delta s)^{\mathrm{lat}}(a)\,,\\
a_0=\Delta\Sigma^{\overline{\mathrm{MS}}}(\mu)&=(\Delta u +\Delta d +\Delta s)^{\overline{\mathrm{MS}}}(\mu)\nonumber\\\label{eq:dq1}
&=Z^{\mathrm{s}}_A(\mu,a)(\Delta u +\Delta d +\Delta s)^{\mathrm{lat}}(a)\,.
\end{align}
We remark that for non-equal quark masses
the non-singlet
combinations Eqs.~(\ref{eq:dq3}) and (\ref{eq:dq2})
also receive contributions from disconnected
quark line diagrams.

\begin{figure}[t]
\centerline{
\includegraphics[height=.43\textwidth,angle=270,clip=]{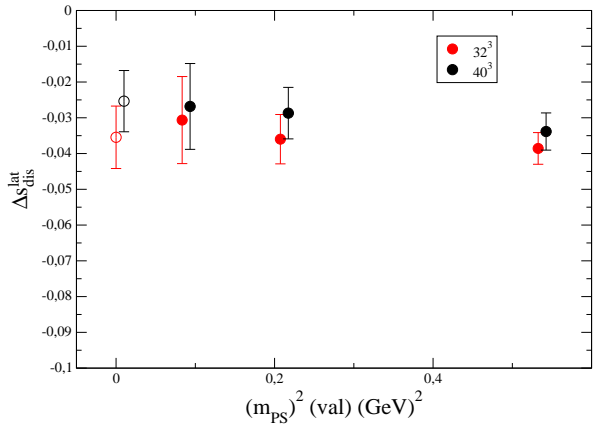}}
\caption{Volume and valence quark mass dependence
of the unrenormalized $\Delta s^{\mathrm{lat}}$.\label{fig:deltas}}
\end{figure}

We employ $n_{\mathrm{f}}=2$ sea quarks
so that our singlet current is $\Delta u +\Delta d$ rather than
the $\Delta\Sigma$ of Eq.~(\ref{eq:dq1}).
This modifies the renormalization
pattern:
\begin{widetext}
\begin{equation}\label{eq:dqren}
\left(
\begin{array}{c}
\Delta u(\mu)\\
\Delta d(\mu)\\
\Delta s(\mu)\end{array}\right)^{\overline{\mathrm{MS}}}
=
\left(
\begin{array}{ccc}
Z_A^{\mathrm{ns}}(a)+\frac{z(\mu,a)}{2}&\frac{z(\mu,a)}{2}&0\\
\frac{z(\mu,a)}{2}&Z_A^{\mathrm{ns}}(a)+\frac{z(\mu,a)}{2}&0\\
\frac{z(\mu,a)}{2}&\frac{z(\mu,a)}{2}&Z_A^{\mathrm{ns}}(a)\end{array}\right)
\left(
\begin{array}{c}
\Delta u(a)\\
\Delta d(a)\\
\Delta s(a)\end{array}\right)^{\mathrm{lat}}\,.
\end{equation}
\end{widetext}
$\Delta s^{\overline{\mathrm{MS}}}$ receives light-quark
contributions but the $\Delta u^{\overline{\mathrm{MS}}}$ and $\Delta d^{\overline{\mathrm{MS}}}$
remain unaffected by the (quenched) strange quark.
Obviously, unitarity is violated, due to this quenching.
The combination $\Delta T_8$ still transforms
with $Z_A^{\mathrm{ns}}$ [Eq.~(\ref{eq:dq2})] while
Eq.~(\ref{eq:dq1}) is violated, as it should be; instead, the $n_{\mathrm{f}}=2$
singlet operator $\Delta u +\Delta d$ renormalizes with $Z_A^{\mathrm{s}}$.
We remark that the above renormalization pattern
is similar to that of the scalar matrix element in the
$n_{\mathrm{f}}=2$ theory~\cite{Gockeler:2004rp,arXiv:1111.1600,Babich:2010at}.
Note that in spite of the quenched strange quark
the mismatch between directly converting the result
into the $\overline{\mathrm{MS}}$ scheme at a scale
$\mu$, using $z(n_{\rm f}=2)/2$, and first converting into
the $\overline{\mathrm{MS}}$ scheme at another scale $\mu'$
and subsequently running within the $\overline{\mathrm{MS}}$ scheme with
$\ln(\mu/\mu')\gamma^{\mathrm{s}}_A(n_{\rm f}=3)/3$ to the scale $\mu$
is tiny.

\textit{Results and systematics.}---In Fig.~\ref{fig:deltas} we display the volume and
(light) valence quark mass dependence of our
unrenormalized $\Delta s^{\mathrm{lat}}$.
There are no statistically significant finite size or mass effects.

\begin{table}
\caption{The connected and disconnected 
contributions to $\Delta q^{\mathrm{lat}}$
as well as the renormalized spin content at a scale
$\mu= \sqrt{7.4}$~GeV. (The $\Delta T_i$ are scale-independent.)
The first error is statistical,
the second is from the renormalization. In addition an overall 20~\% systematic
error needs to be added.
\label{tab:spin}}
\begin{center}
\begin{ruledtabular}
\begin{tabular}{ccccc}
$q$&$V, L$&$\Delta q^{\mathrm{lat}}_{\rm con}$&$\Delta q^{\mathrm{lat}}_{\rm dis}$&$\Delta q^{\mathrm{\overline{\mathrm{MS}}}}(\mu)$\\\hline
$u$&        &~1.065(22)&-0.034(16)&~0.794(21)(2)\\
$d$&        &-0.344(14)&-0.034(16)&-0.289(16)(1)\\
$s$&$V=32^364$& 0        &-0.031(12)&-0.023(10)(1)\\
$T_3$&$L\approx2.33\,$fm&~1.409(24)& 0        &~1.082(18)(2)\\
$T_8$&      &~0.721(26)&-0.006(18)&~0.550(24)(1)\\
$\Sigma$&   &~0.721(26)&-0.098(42)&~0.482(38)(2)\\\hline
$u$&        &~1.071(15)&-0.049(17)&~0.787(18)(2)\\
$d$&        &-0.369( 9)&-0.049(17)&-0.319(15)(1)\\
$s$&$V=40^364$& 0        &-0.027(12)&-0.020(10)(1)\\
$T_3$&$L\approx2.91\,$fm&~1.439(17)& 0        &~1.105(13)(2)\\
$T_8$&      &~0.702(18)&-0.044(19)&~0.507(20)(1)\\
$\Sigma$&   &~0.702(18)&-0.124(44)&~0.448(37)(2)
\end{tabular}
\end{ruledtabular}
\end{center}
\end{table}

Using Eqs.~(\ref{eq:dqren}) and (\ref{eq:dqa})
we can renormalize
\begin{equation}\label{eq:reno}
\Delta q^{\overline{\mathrm{MS}}}(\mu)=Z_A^{\mathrm{ns}}(1+b_Aam_q)\Delta q^{\mathrm{lat}}
+\frac{z(\mu)}{2}(\Delta u+\Delta d)^{\mathrm{lat}}
\end{equation}
for $q\in\{u,d,s\}$. As discussed above,
we omit the $\mathcal{O}(a)$ improvement factor
$(b_A^{\mathrm{s}}Z_A^{\mathrm{s}}-b_AZ_A^{\mathrm{ns}})am_{ud}$ of
the $(\Delta u +\Delta d)^{\mathrm{lat}}$ term.
This is of
$\mathcal{O}(\alpha_s^2am_{ud})$ and
numerically negligible.
We display the bare lattice numbers for
the connected and disconnected contributions
to the proton spin and the renormalized
$\mathcal{O}(a)$ improved values in Table~\ref{tab:spin},
for the two volumes.
The $\Delta u^{\overline{\mathrm{MS}}}$ and $\Delta d^{\overline{\mathrm{MS}}}$
values are
reduced by about 0.035, due to the sea
quark contributions while $\Delta s^{\overline{\mathrm{MS}}}$ increases
by 0.002 ($<10$~\%), due to the mixing with light-quark
flavours.

The uncertainties associated to the renormalization are much smaller
than the statistical errors. Below we will only quote
large volume results, with statistical and renormalization
errors added in quadrature.
Error sources that have so far not been accounted for
are the missing continuum limit
extrapolation, the quenching of the strange quark and
simulating at a light sea quark mass value that is four times bigger
than the physical one. There are no indications of radical
quark mass effects: the flavour mixing effects within
the renormalization are small in spite of the comparatively
large $\Delta u$ and $\Delta d$ values.
The dependence on the valence
quark mass is small too; see Fig.~\ref{fig:deltas}.

Nevertheless, having simulated only at one lattice spacing and
sea quark mass, we cannot extrapolate our results
to the physical point.
Consequently, we underestimate the value~\cite{arXiv:0812.3535}
$g_A=1.2670(35)$ from
neutron $\beta$-decays by $13~\%$ and
find $\Delta T_3=1.105(13)$ instead.
Our prediction
$\Delta T_8=0.507(20)$ differs by the same $13~\%$ from the phenomenological
estimate~\cite{arXiv:0812.3535} $a_8=0.585(25)$.
We take this as an indication of the size of the remaining
systematics and add an additional 20~\% error
to all our results.

\textit{Conclusions.}---We determined
the first moments of proton flavour singlet and non-singlet polarized
parton distributions from $n_{\mathrm{f}}=2$ lattice QCD, at a pion
mass of 285 MeV, at a single lattice spacing $a\approx 0.073$~fm.
We found $\Delta\Sigma=\Delta u+\Delta d +
\Delta s=0.45(4)(9)$
and a small negative
$\Delta s=-0.020(10)(4)$, in the $\overline{\mathrm{MS}}$ scheme,
at a scale $\mu=\sqrt{7.4}$~GeV.
We underestimated both $g_A$ and $a_8$ by similar factors
$\approx 0.87$ and this may suggest that some of the systematics cancel
when considering ratios of
matrix elements. Nevertheless, we emphasize that there is a considerable
uncertainty in the $a_8$ value~\cite{hep-ph/9611210} and 
our $\Delta\Sigma$ is already relatively large,
due to the small difference
$\Delta T_8-\Delta\Sigma=-3\Delta s=0.059(29)(12)$.

Interestingly, our results are in remarkable agreement with
the cloudy bag model prediction of~\cite{Bass:2009ed}.
The small (unrenormalized)
$\Delta s^{\mathrm{lat}}$ value obtained recently in~\cite{Babich:2010at}
is also consistent with our study.
Our $\Delta\Sigma$ value is larger than previously expected, however,
it is compatible 
with the latest
COMPASS number~\cite{arXiv:1001.4654}
$a_0(\sqrt{3}\,\mathrm{GeV})=0.35(3)(5)$. The
experimental number may increase further once
smaller $x$-values become accessible.
We suggest relaxing the weak hyperon decay SU(3)$_F$ constraint
on $a_8$ in determinations
of polarized parton distribution
functions~\cite{arXiv:1010.0574,arXiv:1007.0351,arXiv:0904.3821},
and including our $\Delta s$ prediction instead.

\begin{acknowledgments}
\textit{Acknowledgments.}---
This work is supported by the EU
(Grant No.\ 238353, ITN STRONGnet)
and by the DFG Grant No.\ SFB/TR 55.
S.C.\
acknowledges support from the Claussen-Simon-Foundation (Stifterverband
f\"ur die Deutsche Wissenschaft) and
J.Z.\ from the Australian Research Council
(Grant No.\ FT100100005).
Computations were performed on the
SFB/TR55 QPACE supercomputers,
the BlueGene/P (JuGene) and the Nehalem Cluster (JuRoPA) of the
J\"ulich Supercomputer Center, the
IBM BlueGene/L at the EPCC (Edinburgh),
the SGI Altix ICE machines at HLRN (Berlin/Hannover)
and Regensburg's Athene HPC cluster.
The {\sc Chroma} software suite~\cite{Edwards:2004sx} was used and
gauge configurations were generated with the {\sc BQCD} code~\cite{Nakamura:2010qh}.
\end{acknowledgments}

\end{document}